\documentclass[12pt, draftclsnofoot,journal,onecolumn]{IEEEtran}

\usepackage{amsfonts}
\usepackage{cite}
\usepackage{graphicx}
\usepackage{psfrag}
\usepackage{amsmath}
\usepackage{times}
\usepackage[dvips]{epsfig}
\usepackage{amssymb}
\usepackage{subfigure}
\usepackage[center,small]{caption2}
\usepackage[below]{placeins}

\usepackage{url}
\usepackage{stfloats}

\usepackage{array}

\begin{document}

\title{Superimposed XOR: Approaching Capacity Bounds of the Two-Way Relay Channels}

\author{Jianquan Liu,
        Meixia Tao,
        Youyun Xu,
        and Xiaodong Wang
\thanks{This work was partly presented in IEEE GLOBECOM 2009~\cite{GC09:We}.}
\thanks{Jianquan Liu, Meixia Tao and Youyun Xu are with the Department of Electronic Engineering, Shanghai Jiao Tong University, Shanghai, 200240, P. R. China (e-mails: \{jianquanliu, mxtao, xuyouyun\}@sjtu.edu.cn).}
\thanks{Xiaodong Wang is with the Department of Electrical Engineering, Columbia University, New York, NY 10027, USA (e-mail: wangx@ee.columbia.edu).}}

\maketitle

\IEEEpeerreviewmaketitle

\begin{abstract}

In two-way relay channels, bitwise XOR and symbol-level
superposition coding are two popular network-coding based relaying
schemes. However, neither of them can approach the capacity bound
when the channels in the broadcast phase are asymmetric. In this
paper, we present a new physical layer network coding (PLNC) scheme,
called \emph{superimposed XOR}. The new scheme advances the existing
schemes by specifically taking into account the channel asymmetry as
well as information asymmetry in the broadcast phase. We obtain its
achievable rate regions over Gaussian channels when integrated with
two known time control protocols in two-way relaying.
We also demonstrate their average maximum sum-rates and service
delay performances over fading channels.
Numerical results show that the proposed superimposed XOR achieves a
larger rate region than both XOR and superposition
and performs much better over fading channels.
We further deduce the boundary of its achievable rate region of the
broadcast phase in an explicit and analytical expression. Based on
these results, we then show that the gap to the capacity bound
approaches zero at high signal-to-noise ratio.

\end{abstract}



\section{Introduction}

Cooperative communications enables different users in a wireless
network to share their antennas and cooperate in signal transmission
at the physical layer. This opens up the possibilities of exploiting
distributed spatial diversity and hence effectively enhancing the
system performance. Cooperative communications has thus attracted
significant amount of interests from both academia and industry with
applications in ad-hoc as well as infrastructure-based network. A
basic building block of a cooperative network is the relay channel,
first proposed by van der Meulen~\cite{AAP71:Meulen} and then
extensively studied from information theoretic perspectives by Cover
and Gamal~\cite{IT79:Cover}. The classic relay channel consists of
three nodes, wherein a source node communicates with a destination
node with the help of a relay node. Thus far, a number of relay
schemes have been proposed. Among them, three popular strategies are
known as amplify-and-forward (AF), decode-and-forward (DF) and
compress-and-forward (CF) respectively. However, due to the
half-duplex constraint arising from practical considerations, these
traditional relay schemes suffer from loss in spectral efficiency.

Two-way relaying, where two source nodes exchange information with
the help of a relay node, has recently gained a lot of research
interests~\cite{CISS05:Wu, ADHOC05:Larsson, VTC06:Popovski,
ICC06:Hausl, AMC06:Zhang, ASC06:Katti, JSAC07:Rankov}. It is shown
able to overcome the half-duplex constraint and significantly
improve the system spectral efficiency in relay-based cooperative
networks. Upon receiving the bidirectional information flows, the
relay node combines them together and then broadcasts to the two
desired destinations. The operation at the relay resembles network
coding~\cite{IT00:Ahlswede}, a technique originally developped for
wireline networks. It is thus often referred to as physical layer
network coding (PLNC)~\cite{AMC06:Zhang} or analog network coding
(ANC)~\cite{ASC06:Katti}.

Researchers have attempted to find the best achievable rate region
of two-way relay channels~\cite{IT08:Oechtering, CWIT07:Xie,
IT08:Kim}. Oechtering, {\it et al.} obtained the capacity region of
the broadcast phase in terms of the maximal probability of
error~\cite{IT08:Oechtering}. The achievable rate region of a
two-way relay channel considering both the multiple-access phase
(i.e. the two source nodes transmit simultaneously to the relay
node) and the broadcast phase was studied by Xie using random
binning~\cite{CWIT07:Xie}. Kim, {\it et al.}~\cite{IT08:Kim,
ISS08:Kim} further broadened the frontier of the achievable rate
region by allowing time share between different transmission phases.
In particular, the capacity region of a two-way relay channel with
two-step time protocol is now well-known to be the intersection of
the optimal time-weighted capacity regions of the multiple-access
phase and the broadcast phase. Note that the above capacity analysis
all assumed full decoding at the relay\footnote{The capacity region
of two-way relaying with partial decoding still remains open.}.
Moreover, the information theoretic techniques including random
binning and jointly typical set decoding are often adopted at the
relay and destinations in the broadcast phase~\cite{IT08:Oechtering,
CWIT07:Xie, IT08:Kim}.

Meanwhile, a number of practical PLNC schemes for two-way relay
channels have also been proposed and analyzed, such as bit-level
XOR~\cite{ICC07:Popovski, ICC08:Liu}, symbol-level superposition
coding~\cite{JSAC07:Rankov, CL08:Oechtering} and
AF~\cite{AMC06:Zhang, JSAC07:Rankov}. In particular, authors
in~\cite{JSAC07:Rankov} obtained the rate pair expressions for
superposition based relaying. Authors in~\cite{ICC07:Popovski}
analyzed the maximum achievable sum-rate for XOR based relaying with
optimal time control. It is worth mentioning that the asymmetry in
both packet size and channel gain of the two transmitting nodes are
considered in~\cite{ICC07:Popovski}. Oechtering also studied the
optimal time control for superposition
scheme~\cite{CL08:Oechtering}. Despite all these attempts, there is
still a large gap between the rate regions achieved by the practical
PLNC schemes and the capacity bound in asymmetric relay channels, as
shown by a recent comparative study in~\cite{CC09:We}.

In this paper, we study practical and capacity approaching PLNC
schemes over the two-way relay channels. In this regard, we propose
a novel PLNC scheme, named as \emph{superimposed XOR}, tailored for
the broadcast phase of two-way relaying with asymmetric channels.
Combining it with two known transmission protocols: 4-step with
direct link~\cite{IT08:Kim} and 2-step with no direct link, we
analyze its achievable rate regions over Gaussian channels.
We also demonstrate their average maximum sum-rates and service
delay performances over fading channels.
Numerical results show that the performance of the proposed
superimposed XOR outperforms the traditional XOR and superposition
in terms of achievable rate region. It also closely approaches the
optimal capacity bounds of the two-way relay channels in the high
signal-to-noise ratio (SNR) regime. To further illustrate the
capacity approaching behavior of the proposed superimposed XOR, we
obtain the analytical expressions of the boundary of its achievable
broadcast rate region. Based on these results, we then explicitly
prove that its gap with the capacity of the broadcast phase
approaches zero when the SNR is much larger than one.

The rest of the paper is organized as follows. In Section II, we
present the system model of two-way relaying. In Section III, we
describe the proposed superimposed XOR scheme. Section IV
characterizes the rate regions over Gaussian channels. The capacity
approaching performance is analyzed in Section V. Finally, we
conclude the paper in section VI.

\section{System Model}

We consider a two-way relay channel which consists of two source
nodes and one relay node. The source nodes, denoted as 0 and 2, wish
to exchange information with the help of the relay node, denoted as
1. We assume that all the nodes operate in the half-duplex mode. The
channel on each communication link is assumed to be corrupted with
Rayleigh fading and additive white Gaussian noise. The SNR of the
link from node $i$ to node $j$ is denoted as $\gamma_{ij}$, for $i,j
\in \{0,1,2\}$, and it counts both channel gain and transmit power.
Note that $\gamma_{ij}$ may not be equal to $\gamma_{ji}$ as the
channels considered here may not be reciprocal. The channel capacity
in bit/s/Hz of the link from node $i$ to node $j$ is denoted as
$C_{ij}$, and determined by the SNR on the link as
\begin{eqnarray}
  C_{ij} &\triangleq & \emph{C}(\gamma_{ij}) = \log_2(1 +
\gamma_{ij}).
\end{eqnarray}
Throughout this paper, we assume that the relay needs to fully
decode the information of the two source nodes. The sum-rate
capacity of the multiple access channel (MAC) when nodes $0$ and $2$
are transmitting simultaneously to node $1$ is denoted as
\begin{eqnarray}
C_{{\rm{m}}} \triangleq \emph{C}(\gamma_{01} + \gamma_{21}) =
\log_2(1 + \gamma_{01} + \gamma_{21}).
\end{eqnarray}

\subsection{Time control protocols}

Two-way relaying involves not only PLNC at the relay node but also
time control for node transmission. In this subsection, we review
some existing time control protocols. Similar
to~\cite{ICC07:Popovski}, we name the protocols based on the
number of time steps to finish one round of information exchange
between the source nodes. We focus on the 4-step and 2-step
protocols in this paper.

\subsubsection{4-step protocol}

In the first step, node 0 transmits for $\lambda_1$ time duration
and node 1 and 2 listen. In the second step, node 2 transmits for
$\lambda_2$ time duration and node 1 and 0 listen. In the third
step, nodes 0 and 2 transmit simultaneously for $\lambda_3$ time
duration while node 1 listens. In the fourth step, node 1
transmits for $\lambda_4$ time duration and node 0 and 2 listen.
Without loss of generality, the total time duration is normalized
to one, i.e., $\sum_{i=1}^4\lambda_i = 1$.

\subsubsection{2-step protocol}

In 2-step protocol, nodes 0 and 2 first transmit simultaneously for
$\lambda_1$ time duration while node 1 listens. Next, node 1
transmits for $\lambda_2$ time duration and nodes 0 and 2 listen.
This protocol can be regarded as a special case of the 4-step
protocol by letting $\lambda_1 = \lambda_2 = 0$. No direct link is
exploited here.

\section{Superimposed XOR}

Upon decoding the two bit sequences ${\bf{b}}_0$ and ${\bf{b}}_2$
from the source nodes 0 and 2 to be exchanged, the relay node will
perform physical layer network coding on ${\bf{b}}_0$ and
${\bf{b}}_2$ and then broadcast to the two destinations. Before
introducing the proposed PLNC scheme, we briefly discuss the
bit-level XOR and symbol-level superposition coding which the
proposed scheme is based upon.

\subsection{Bit-level XOR}

The relay node performs bitwise XOR on the two bit sequences as
${\bf{b}}_0 \oplus {\bf{b}}_2$. In the case where the lengths of the
two sequences are not equal, there are two methods to perform the
XOR, as shown in Fig.~\ref{fig:framework}(a). The first one is to
pad the shorter sequence with zero bits so as to make it having the
same length as the longer sequence and then perform XOR. In the
second method, the longer sequence will be partitioned into two
sub-sequences, with one having the same length as the short
sequence. XOR is then performed on the shorter sequence and the
sub-sequence with equal length. The resulting bit sequence is
broadcasted to both receivers. The other sub-sequence will be
transmitted alone, at a possibly higher rate, to its desired
receiver. In practice, which method to use depends on the
relationship of the channel gains in the broadcast phase of two-way
relaying~\cite{ICC07:Popovski}. In general, if the channel quality
of the receiver of the longer sequence is worse than the channel
quality of the receiver of the shorter sequence, zero-padding is
applied Otherwise, partitioning the longer packet is preferred.

\subsection{Symbol-level superposition}
The relay encodes the two bit sequences ${\bf{b}}_0$ and
${\bf{b}}_2$ separately into baseband signal sequences ${\bf x}_0$
and ${\bf x}_2$ with the same length, and then superimposes them
together as $\sqrt{\theta}{\bf x}_0 + \sqrt{1-\theta}{\bf x}_2$, as
shown in Fig.~\ref{fig:framework}(b). Here $\theta$ is a power
allocation coefficient~\cite{JSAC07:Rankov}. The signal
$\sqrt{\theta}{\bf x}_0 + \sqrt{1-\theta}{\bf x}_2$ is then
broadcasted directly without further encoding. Unlike bit-level XOR,
there is no need to consider the issue of asymmetry in bit length.
Another essential difference between superposition and XOR is that,
the information combining is carried out in the symbol level after
channel coding and modulation for the former, while it is in the
information bit level before channel coding for the latter.

\subsection{Proposed superimposed XOR}

As discussed in \cite{CC09:We}, for symmetrical broadcast channels
($\gamma_{10} = \gamma_{12}$), bit-level XOR is capacity-achieving
whereas superposition coding is suboptimal. But for asymmetrical
channels ($\gamma_{10} \ne \gamma_{12}$), bit-level XOR becomes
inferior to superposition at certain rate regions. The proposed
superimposed XOR is specifically designed for asymmetrical broadcast
channels and it utilizes the advantages of both XOR and
superposition schemes. The details are as follows.

\begin{figure}
\centering
\includegraphics[width=4in]{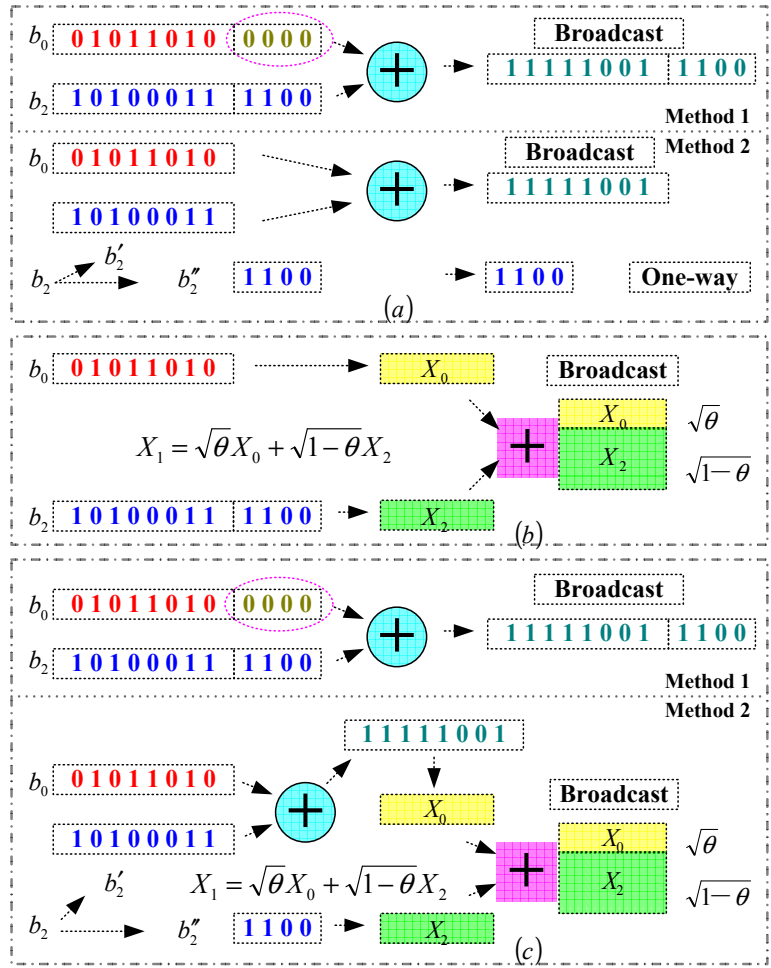}
\caption{Three kinds of PLNC schemes for broadcast phase: (a) XOR,
(b) Superposition, (c) Superimposed XOR.} \label{fig:framework}
\end{figure}

If the lengths of two bit sequences ${\bf{b}}_0$ and ${\bf{b}}_2$
are equal, then the conventional bitwise XOR is performed.
Otherwise, there are two methods to process the two sequences, as
shown in Fig.~\ref{fig:framework}(c). The first method is the same
as the one in bit-level XOR scheme, that is, padding the shorter
sequence, say ${\bf b}_0$, with zero bits, so as to make them equal
and then performing bit-level XOR. The resulting bit sequence, after
channel coding and modulation, is then broadcasted to both
receivers. In the second method, the relay node first partitions the
longer sequence, say ${\bf b}_2$, into two sub-sequences as ${\bf
b}_2 = [{\bf b}_2' {\bf b}_2'']$, where the sub-sequence ${\bf
b}_2'$ has the same length as ${\bf b}_0$. It then encodes the XORed
sub-sequence ${\bf b}_0 \oplus {\bf b}_2'$ and the sub-sequence
${\bf b}_2''$, separately. We denote the resulting coded symbol
sequences as ${\bf x}_0$ and ${\bf x}_2$. Finally, the relay
superimposes them together as $\sqrt{\theta}{\bf x}_0 +
\sqrt{1-\theta}{\bf x}_2$ which is broadcasted directly to the two
destinations. Here, $\theta$ is also a power allocation coefficient,
${\bf x}_0$ is to be received by both destinations, and ${\bf x}_2$
is to be received by one destination only. Which of the above two
methods to use depends on the relationship between $\gamma_{10}$ and
$\gamma_{12}$ in the broadcast channels. Suppose that the longer
sequence is ${\bf b}_2$ which is to be sent to node 0. Then, if
$\gamma_{10} < \gamma_{12}$, we apply method one and else apply
method two. We shall discuss this in more detail in the proof of the
rate regions in the next section.

Note that Liu, {\it et al.} proposed a joint network coding and
superposition coding (JNSC) scheme for information exchange among
more than two users in a wireless relay network~\cite{GC08:Liu}.
Therein, two XOR-ed packets generated by information from three
nodes are superimposed. Our proposed superimposed XOR scheme differs
from the JNSC~\cite{GC08:Liu} in that our scheme performs
superposition on only one XOR-ed packet and the sub-packet obtained
by partitioning the longer bit sequence.
A similar PLNC scheme as the second method of our proposed
superimposed XOR was proposed by Chen, {\it et al.} for multi-hop
wireless networks in~\cite{ICC06:Chen}. The difference however is
that our scheme adaptively selects the aforementioned two methods
illustrated in Fig.~\ref{fig:framework}(c) according to the channel
conditions in the broadcast phase while the scheme
in~\cite{ICC06:Chen} is fixed at method two regardless the channel
conditions. Such static approach could be far from
capacity-achieving as it does not exploit the channel dynamics.

In the rest of the paper, for ease of presentation, if the
considered three PLNC schemes (XOR, superposition, and superimposed
XOR) are combined with the 4-step time control protocol, we denote
them as 4S-XOR, 4S-SUP and 4S-SuX, respectively. Likewise, when
combined with 2-step protocol, they are named as 2S-XOR, 2S-SUP and
2S-SuX.

\section{Analysis of rate regions}

Let $R_0$ and $R_2$ denote the data rates of the information flows
$0 \rightarrow 2$ and $2 \rightarrow 0$, respectively, in the
considered two-way communications. In this section, we derive the
rate region $(R_0, R_2)$ of the aforementioned relay strategies.
Without loss of generality, we assume $\gamma_{10} \geq
\gamma_{12}$ and thus $C_{10} \geq C_{12}$ in Subsections IV-A and
B.

\subsection{Achievable rate region for superimposed XOR}

{\it Theorem 1 (4S-SuX)}: The rate region for 4S-SuX is the
closure of the set of all rate pairs $(R_0, R_2)$ satisfying
\begin{equation*}
\label{eq:4}
\begin{array}{l}
(R_0, R_2): \Big\{R_0 \leq \min\big((\lambda_1 + \lambda_3)C_{01},
\lambda_1C_{02} + \lambda_4C_{12}(\theta)\big),
\\~~~~R_2 \leq \min\big((\lambda_2 + \lambda_3)C_{21}, \lambda_2C_{20} + \lambda_4C_{10}\big),
\\~~~~R_2 - R_0 \leq - \lambda_1C_{02} + \lambda_2C_{20} + \lambda_4C_{10}(1-\theta),
\\~~~~R_0 + R_2 \leq \lambda_1C_{01} +
\lambda_2C_{21} + \lambda_3C_{\rm{m}}, \\
~~~~\sum_{i=1}^4\lambda_i = 1, \theta \in [0,1] \Big\}.
\end{array}
\end{equation*}
Here $C_{ij}(\theta) \triangleq  \emph{C}(\gamma_{ij}\theta) =
\log_2(1 + \gamma_{ij}\theta)$.

\begin{figure}
\centering
\includegraphics[width=4in]{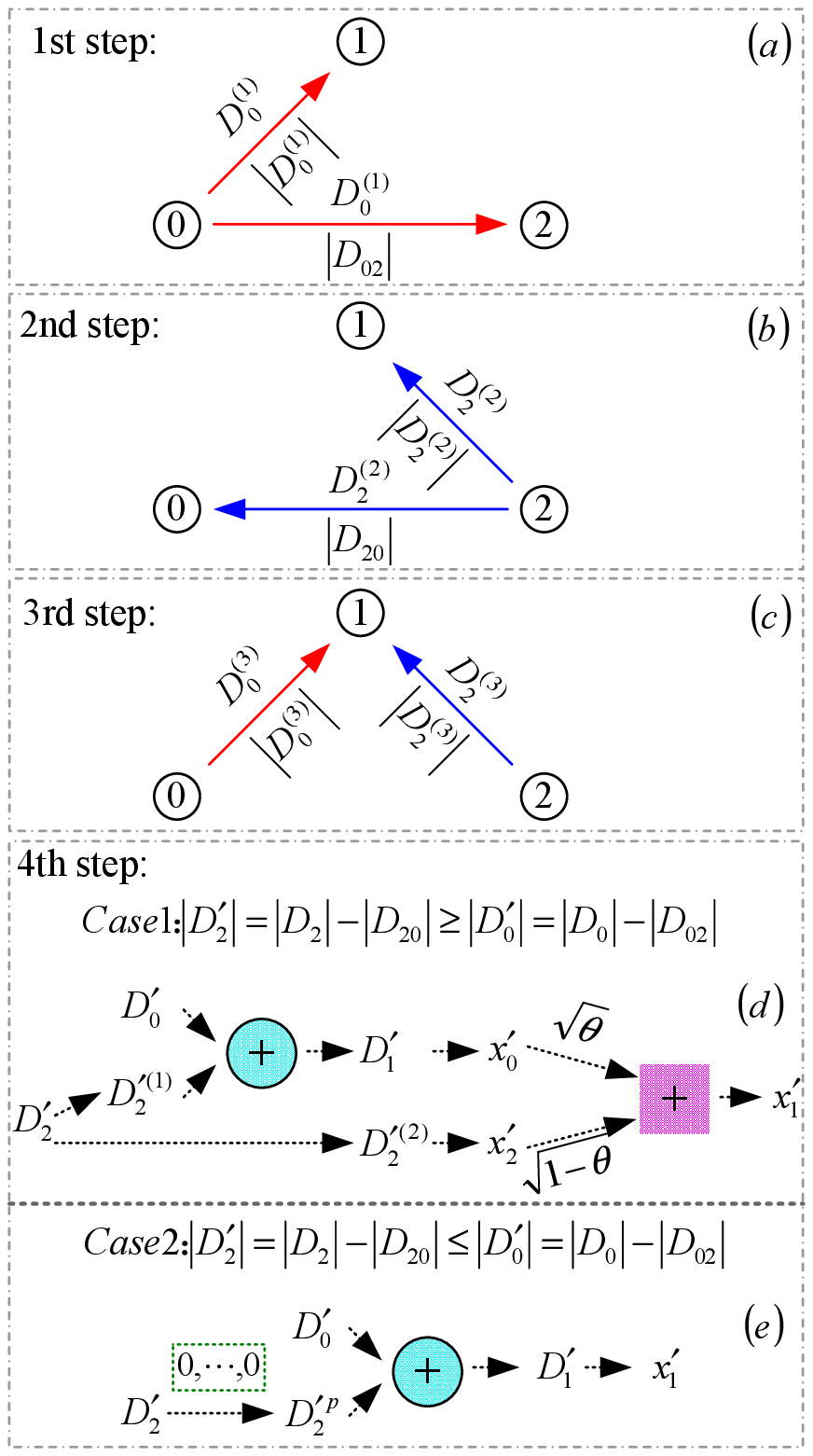}
\caption{Process flow of the 4s-Sux.} \label{fig:proof_aide}
\end{figure}

\begin{proof}
For ease of comprehension, Fig.~\ref{fig:proof_aide} is presented to
assist the proof. Let $D_0$ denote the information packet to be
transmitted from node 0 to 2 and its packet length in bits be
denoted as $|D_0|$. Assume the message in the packet is further
split into two parts, denoted as $D_0^{(1)}$ and $D_0^{(3)}$, which
are transmitted in the first and third step, respectively, as
illustrated in Fig.~\ref{fig:proof_aide} (a) and (c). Likewise, we
let $D_2$ denote the information packet to be transmitted from node
2 to 0, and let it be split into $D_2^{(2)}$ and $D_2^{(3)}$ for
transmission in the second and third steps, as depicted in
Fig.~\ref{fig:proof_aide} (b) and (c). During the first three steps
of packet transmission (i.e. the multiple-access phase), it is
obvious that $|D_{0}^{(1)}| \leq \lambda_1 C_{01}$, $|D_{2}^{(2)}|
\leq \lambda_2 C_{21}$, $|D_{0}^{(3)}| \leq \lambda_3 C_{01}$,
$|D_{2}^{(3)}| \leq \lambda_3 C_{21}$, and $|D_{0}^{(3)}| + |
D_{2}^{(3)}| \leq \lambda_3 C_{\rm{m}}$. Note that in the first
step, due to the presence of direct link between node 0 and node 2,
the desired destination node 2 is able to exact $|D_{02}| \leq
\lambda_1 C_{02}$ amount of information. Thus, the total amount of
information bits to be transmitted through the relay link in the
fourth step to node 2 is $|D_0| - |D_{02}|$ and we denote the
corresponding packet as $D_0'$. Similarly, the total amount of
information to be relayed from node 2 to node 0 in the fourth step
is $|D_2| - |D_{20}|$, with $|D_{20}| \leq \lambda_2 C_{20}$, and
the corresponding packet can be denoted as $D_2'$. Then, during the
fourth step transmission (i.e. the broadcast phase), by comparing
the packet sizes $|D_0'|$ and $|D_2'|$, two cases need to be
considered.

{\it Case 1:} $|D_{2}'| \geq |D_{0}'|$. As shown in
Fig.~\ref{fig:proof_aide} (d), the relay node 1 partitions the
packet $D_{2}'$ into $D_{2}'^{(1)}$ and $D_{2}'^{(2)}$ so that
$|D_{2}'^{(1)}| = |D_{0}'|$ and $|D_{2}'^{(2)}| = |D_{2}'| -
|D_{0}'|$. Packet $D_{2}'^{(1)}$ contains the first $|D_{0}'|$ bits
from $D_{2}'$ and packet $D_{2}'^{(2)}$ contains the rest of the
bits from $D_{2}'$. Now, node 1 creates $D_{1}' = D_{2}'^{(1)}
\oplus D_{0}'$. Then, the information bits $D_{1}'$ and
$D_{2}'^{(2)}$ are encoded separately into two codewords ${\bf
x}_0'$ and ${\bf x}_2'$ with the same length, which are then
superimposed together in the complex field. Unlike XOR-based scheme,
there is no extra time used to transmit $D_{2}'^{(2)}$ to node 0.
Let $\theta$ present the power ratio allocated to the signal ${\bf
x}_0'$ to be transmitted to nodes 0 and 2 and $1-\theta$ be the
power ratio on the signal ${\bf x}_2'$ to node 0, where $0 \leq
\theta \leq 1$. Since $D_{2}'$ is known at node 2 and so is ${\bf
x}_2'$, to successfully decode the packet $D_{1}'$ at node 2, the
transmission rate of ${\bf x}_0'$ in the fourth step with a fraction
$\lambda_{4}$ of time cannot exceed
$C_{12}(\theta)=\emph{C}(\theta\gamma_{12})$. Thus, we have $|D_{0}|
- |D_{02}| = |D_{0}'| = |D_{1}'| \leq \lambda_{4}C_{12}(\theta)$.
Since node 0 does not know the packets $D_{2}'^{(1)}$ and
$D_{2}'^{(2)}$ and only node 0 needs to decode ${\bf x}_2'$, the
link $1 \to 0$ can be regarded as a virtual multiple-access channel
(MAC) and the channel capacity is bound by $|D_{2}'^{(1)}| =
|D_{1}'| \leq \lambda_4C_{12}(\theta)$, $ |D_2'| - |D_{0}'| = |D_2'|
- |D_{2}'^{(1)}| = |D_{2}'^{(2)}| \leq \lambda_4C_{10}(1-\theta)$
and $|D_{2}'| = |D_{2}'^{(1)}| + |D_{2}'^{(2)}| \leq
\lambda_4\emph{C}(\theta\gamma_{10} + (1-\theta)\gamma_{10}) =
\lambda_4C_{10}$.

After receiving ${\bf x}_1 = \sqrt{\theta}{\bf x}_0' +
\sqrt{1-\theta}{\bf x}_2'$, node 2 extracts the symbol ${\bf x}_0'$
as ${\bf x}_0' = {\bf x}_1 - \sqrt{1-\theta}{\bf x}_2'$, where ${\bf
x}_2'$ is encoded by $D_{2}'^{(2)}$ which already has been known. By
decoding ${\bf x}_0'$, we get $D_{1}'$. Then, node 2 extracts the
packet $D_{0}'$ as $D_{0}' = D_{1}' \oplus D_{2}'^{(1)}$. Similarly,
after receiving ${\bf x}_1$, node 0 extracts ${\bf x}_0'$ and ${\bf
x}_2'$ by fully decoding. Then, $D_{1}'$ and $D_{2}'^{(2)}$ can be
obtained easily. Next, node 0 extracts the packet $D_{2}'^{(1)}$ as
$D_{2}'^{(1)} = D_{1}' \oplus D_{0}'$. Note that, for each
destination, say node 2, to recover the desired packet $D_0$ from
$D_{02}$ and $D_{0}'$, a coding method for Gaussian parallel channel
should be employed~\cite{IT06:Anders}.

With the constraints obtained from the above discussion and using
the definition that $R_0 = |D_0|$ and $R_2 = |D_2|$, we obtain the
set of linear inequalities about $R_0$ and $R_2$ after simple
manipulation:
\setlength\arraycolsep{2pt}
\begin{eqnarray*}
\label{eq:4} R_0 &\leq & \min\big((\lambda_1 + \lambda_3)C_{01},
\lambda_1C_{02} + \lambda_4C_{12}(\theta)\big), \\
R_2 &\leq& \min\big((\lambda_2 + \lambda_3)C_{21}, \lambda_2C_{20}
+ \lambda_4C_{10}\big),\\
R_2 - R_0 &\leq &- \lambda_1C_{02} + \lambda_2C_{20} +
\lambda_4C_{10}(1-\theta), \\
R_0 + R_2 &\leq &\lambda_1C_{01} + \lambda_2C_{21} +
\lambda_3C_{\rm{m}},\\
R_0 - R_2 &\leq & \lambda_1C_{02} - \lambda_2C_{20}.
%
\end{eqnarray*}

In addition, due to the total time constraint, we have
$\sum_{i=1}^4 \lambda_i = 1$. Lastly, the power ratio $\theta$ can
take any value that satisfies $0\le \theta \le 1$.

{\it Case 2:} $|D_{2}'| \leq |D_{0}'|$. As in
Fig.~\ref{fig:proof_aide} (e), the packet $D_{2}'$ is padded with
zeros to obtain the packet $D_{2}'^{p}$ such that $|D_{2}'^{p}| =
|D_{0}'|$. Since node 0 and 2 know the size of $D_{2}'$, they also
know how many zeros are used for padding. Node 1 creates the packet
$D_{1}' = D_{2}'^{p} \oplus D_{0}'$. In Step 4 the packet $D_{1}'$
is broadcasted at a rate at which both node 0 and node 2 can
successfully decode. Thus, we have $R_0 - R_{02} = |D_{0}| -
|D_{02}| = |D_{0}'| = |D_{1}'| \leq \lambda_4 C_{12}$. Node 2 then
extracts $D_{0}'$ as $D_{0}' = D_{2}'^{p} \oplus D_{1}'$, which is
the desired packet sent from node 0. Similarly, node 0 can obtain
$D_{2}'^{p}$ from $D_{1}'$. The packet $D_{2}'$ is then obtained by
removing the padding zeros from $D_{2}'^p$.

Thus, in this case, we obtain the following linear inequalities
that the rate pair $(R_0, R_2)$ has to satisfy:
\begin{eqnarray*}
R_0 &\leq & \min\big((\lambda_1 + \lambda_3)C_{01},
\lambda_1C_{02} + \lambda_4C_{12}\big),\\
R_2 &\leq & (\lambda_2 + \lambda_3)C_{21}, \\
R_0 + R_2 &\leq & \lambda_1C_{01} + \lambda_2C_{21} +
\lambda_3C_{\rm{m}},\\
R_2 - R_0 &\leq &- \lambda_1C_{02} + \lambda_2C_{20}.
\end{eqnarray*}

Finally, combining the set of results for case 2 with the results
for case 1, we obtain the rate region of 4S-SuX as given in the
theorem.
\end{proof}

{\it Remark}: By setting $\lambda_1 = \lambda_2 = 0$ in Theorem 1,
we obtain the rate region for 2S-SuX.

{\it Theorem 2 (2S-SuX)}: The rate region for 2S-SuX is the
closure of the set of all rate pairs $(R_0, R_2)$ satisfying
$$\label{eq:4}
\begin{array}{l}
(R_0, R_2):\Big\{R_0 \leq \min\big(\lambda_1C_{01},
\lambda_2C_{12}(\theta)\big),
\\~~~~R_2 \leq \min(\lambda_1C_{21}, \lambda_2C_{10}),
\\~~~~R_2 - R_0 \leq \lambda_2C_{10}(1-\theta),
\\~~~~R_0 + R_2 \leq \lambda_1C_{\rm{m}},
\\~~~~\sum_{i=1}^2\lambda_i = 1, \theta \in [0,1] \Big\}.
\end{array}
$$

\subsection{Achievable rate region for XOR}

{\it Theorem 3 (4S-XOR)}: The rate region for 4S-XOR is the
closure of the set of all rate pair $(R_0, R_2)$ satisfying
$$\label{eq:4}
\begin{array}{l}
(R_0,R_2):\Big\{R_0 \leq \min\big((\lambda_1 + \lambda_3)C_{01}, \lambda_1C_{02} + \lambda_{4}C_{12}\big),\\
R_0 + R_2 \leq \lambda_1C_{01} + \lambda_2C_{21} + \lambda_3C_{\rm{m}}, R_2 \leq (\lambda_2 + \lambda_3)C_{21},\\
R_2 - R_0 \leq -\lambda_1C_{02} + \lambda_2C_{20} +
\lambda_{5}C_{10}, \sum_{i=1}^5\lambda_i = 1\Big\}.
\end{array}
$$

\begin{proof}
The proof of this theorem differs from the proof of Theorem 1 mainly
in Case 1 of the broadcast phase. The coding and decoding method of
Case 1 are similar to those discussed in~\cite{ICC07:Popovski,
ICC08:Liu} and are omitted.

\end{proof}

{\it Remark}: A point to note is that five time-sharing parameters
are needed when the 4-step time protocol is combined with XOR. This
is because the last step (broadcast phase) in the XOR scheme can be
further divided into two sub-steps, if necessary, according to the
discussion in Section III-A.

{\it Remark}: By setting $\lambda_1 = \lambda_2 = 0$ in Theorem 3,
we obtain the rate region for 2S-XOR.

{\it Theorem 4 (2S-XOR)}: The rate region for 2S-XOR is the
closure of the set of all rate pair $(R_0, R_2)$ satisfying
$$\label{eq:4}
\begin{array}{l}
(R_0, R_2):\Big\{R_0 \leq \min\big(\lambda_1 C_{01}, \lambda_{2}C_{12}\big), R_2 \leq \lambda_1 C_{21},\\
R_0 + R_2 \leq \lambda_1 C_{\rm{m}}, R_2 - R_0 \leq \lambda_{3}
C_{10}, \sum_{i=1}^3\lambda_i = 1 \Big\}.
\end{array}
$$

Note that the authors in \cite{ICC07:Popovski} also studied the
rate pair of 2S-XOR, but only the maximum sum-rate is considered
and no rate region is discussed. In addition, the work in
\cite{ICC07:Popovski} is only suitable for reciprocal channels
with $C_{01}=C_{10}$ and $C_{21}=C_{12}$.

\subsection{Achievable rate region for superposition coding}

The achievable rate region of 2S-SUP is well studied
in~\cite{JSAC07:Rankov, CL08:Oechtering, TC08:Oechtering}. Though
the achieve rate region of 4S-SUP is not studied yet in the
literature, its derivation is trivial.

{\it Theorem 5 (4S-SUP)}: The rate region for 4S-SUP is the
closure of the set of all rate pair $(R_0, R_2)$ satisfying
$$\label{eq:4}
\begin{array}{l}
(R_0, R_2):\Big\{R_0 \leq \min\big((\lambda_1 + \lambda_3)C_{01},
\lambda_1C_{02} +
\lambda_4C_{12}(\theta)\big),\\
R_2 \leq \min\big((\lambda_2 + \lambda_3)C_{21}, \lambda_2C_{20} +
\lambda_4C_{10}(1-\theta)\big),\\
R_0 + R_2 \leq \lambda_1C_{01} + \lambda_2C_{21} +
\lambda_3C_{\rm{m}},\sum_{i=1}^4\lambda_i =1,\theta \in [0,1]
\Big\}.
\end{array}
$$

{\it Theorem 6 (2S-SUP)}: The rate region for 2S-SUP is the
closure of the set of all rate pair $(R_0, R_2)$ satisfying
$$\label{eq:4}
\begin{array}{l}
(R_0, R_2):\Big\{R_0 \leq \min\big(\lambda_1 C_{01},
\lambda_2C_{12}(\theta)\big), R_0 + R_2 \leq \lambda_1 C_{\rm{m}},\\
R_2 \leq \min\big(\lambda_1 C_{21},
\lambda_2C_{10}(1-\theta)\big), \sum_{i=1}^2\lambda_i = 1, \theta
\in [0,1] \Big\}.
\end{array}
$$

\subsection{Cases with direct transmission}
For 4-step strategies, it is assumed by default that the direct
link is always worse than the relay link. That is,  $C_{02} <
C_{01}$ and $C_{20} < C_{21}$. In this subsection, we consider the
special cases where the direct link is better.

If $C_{02} < C_{01}$ and $C_{20} \geq C_{21}$, the signal from
node 2 will be transmitted directly to node 0 without the help of
relay. This corresponds to $\lambda_3 = 0$ and the relay
transmitting to node 2 only during $\lambda_4$ in the 4-step
protocol. Thus, using Theorem 1, we obtain the sets of rate pairs
$(R_0, R_2)$ satisfying
$$\label{eq:5}
\begin{array}{l}
(R_0, R_2):\Big\{ R_0 \leq \min\big(\lambda_1 C_{01}, \lambda_1
C_{02} + \lambda_2 C_{12}\big), \\R_2 \leq \lambda_3 C_{20},
\sum_{i=1}^3\lambda_i = 1 \Big\}.
\end{array}
$$

If $C_{02} \geq C_{01}$ and $C_{20} < C_{21}$, the signal from
node 0 will be transmitted directly to node 2 without the help of
relay. The sets of rate pairs $(R_0, R_2)$ satisfy
$$\label{eq:6}
\begin{array}{l}
(R_0, R_2):\Big\{R_2 \leq \min\big(\lambda_2 C_{21}, \lambda_2
C_{20} + \lambda_3 C_{10}\big), \\R_0 \leq \lambda_1 C_{01},
\sum_{i=1}^3\lambda_i = 1 \Big\}.
\end{array}
$$

If $C_{02} \geq C_{01}$ and $C_{20} \geq C_{21}$, direct
communication between node 0 and node 2 in both ways is preferred
and no relay is needed. Hence, the rate pairs are given by
$$\label{eq:7}
\begin{array}{l}
(R_0, R_2):\Big\{ R_0 \leq \lambda_1 C_{02}, R_2 \leq \lambda_2
C_{20}, \sum_{i=1}^2\lambda_i = 1 \Big\}.
\end{array}\vspace{-6pt}
$$

\subsection{Numerical results}
In this subsection, we present a numerical study of the proposed
superimposed XOR in terms of three performance metrics: rate
regions, system average sum-rates and service delay performances.

\begin{figure}
\centering
\includegraphics[width=4in]{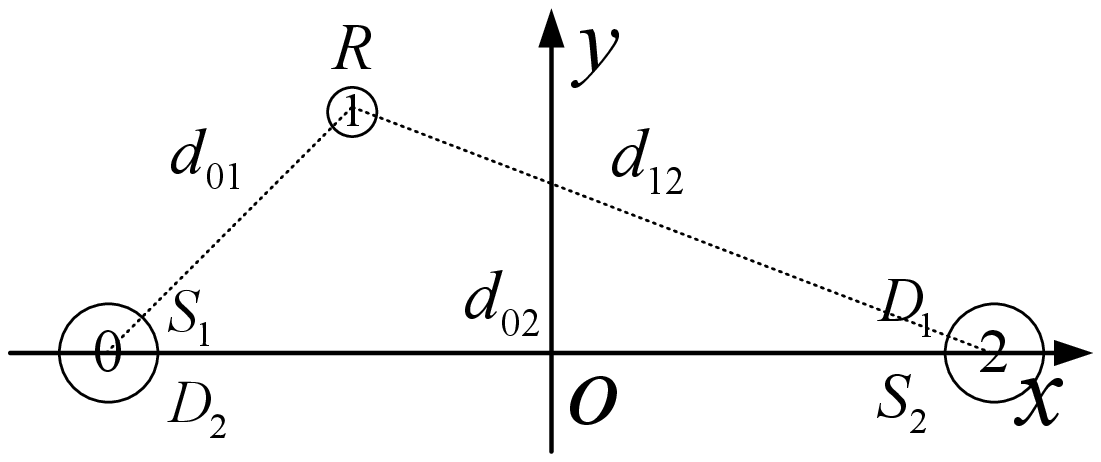}
\caption{Layout of the two-way relaying network} \label{fig:layout}
\end{figure}

Suppose that the channel gain on each link is modeled by the
distance path loss model, given by $\alpha_{ij} = c \cdot
d_{ij}^{-n}$, where $c$ is an attenuation constant, $n$ is the path
loss exponent and fixed at 3, and $d_{ij}$ denotes the distance
between nodes $i$ and $j$. For simplicity, each node uses the same
transmission power $P$, though our analytical results are suitable
for the case with unequal transmit power. The noise power is assumed
to one. We consider the network layout shown in
Fig.~\ref{fig:layout}, where the distance between nodes 0 and 2 is
normalized to 1 and the location of the relay is determined using
the projections $x$ and $y$. The source nodes 0 and 2 are located at
coordinates $(-0.5, 0)$ and $(0.5, 0)$, respectively. The distances
from the relay to the source nodes can be computed as
$d_{01}=\sqrt{(x+0.5)^2+y^2}$, and $d_{12}=\sqrt{(x-0.5)^2+y^2}$.

For fading channels, the same network layout and channel model,
except that small-scale fading is included. We assume that the
fading on each link follows Rayleigh distribution and are
independent and reciprocal for different links.

\subsubsection{Rate regions in Gaussian channels}

\begin{figure}
\centering
\includegraphics[width=5in]{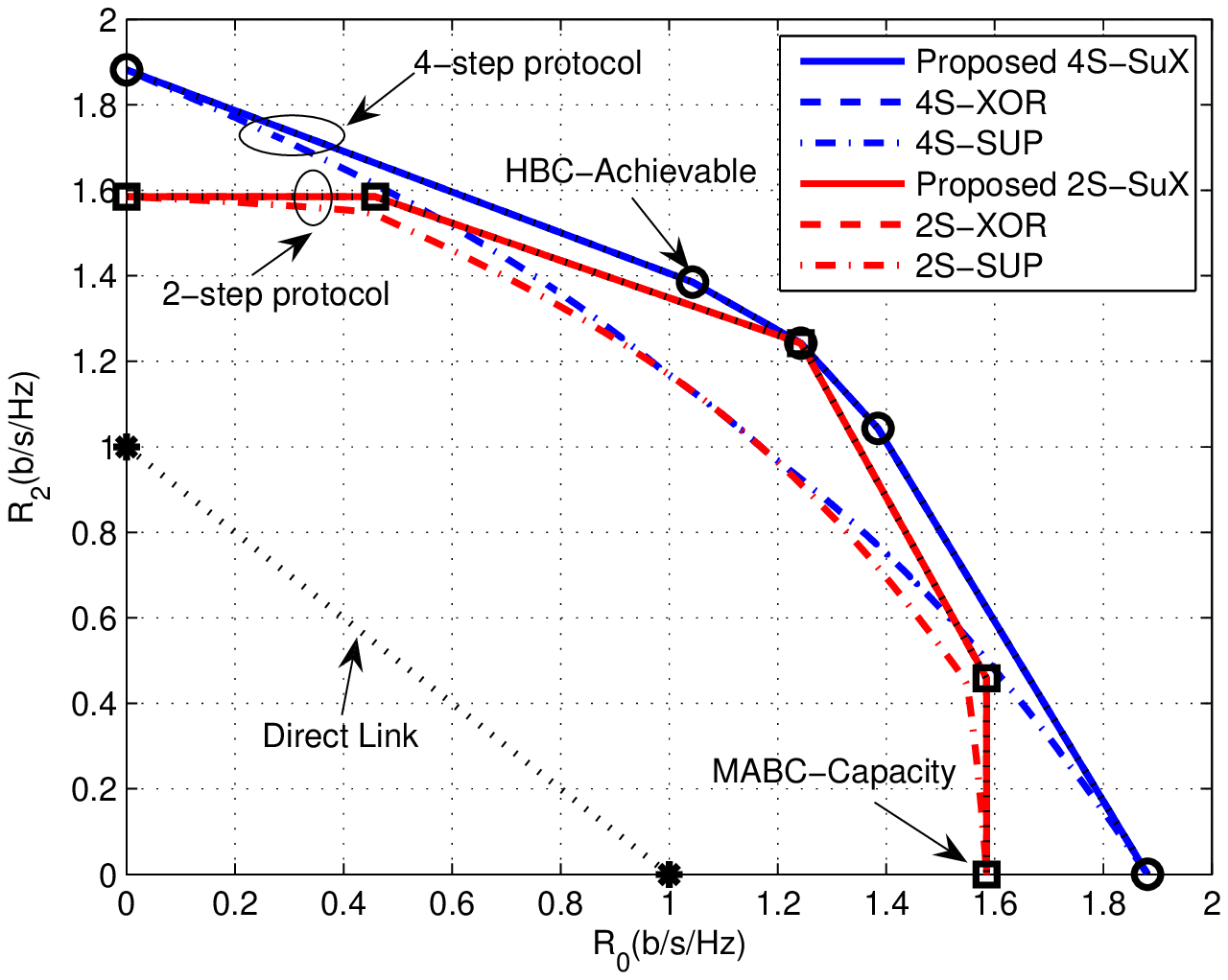}
\caption{Rate region at $(x,y)=(0,0)$ with $P=0$dB.}
\label{fig:region_s}
\end{figure}

\begin{figure}
\centering
\includegraphics[width=5in]{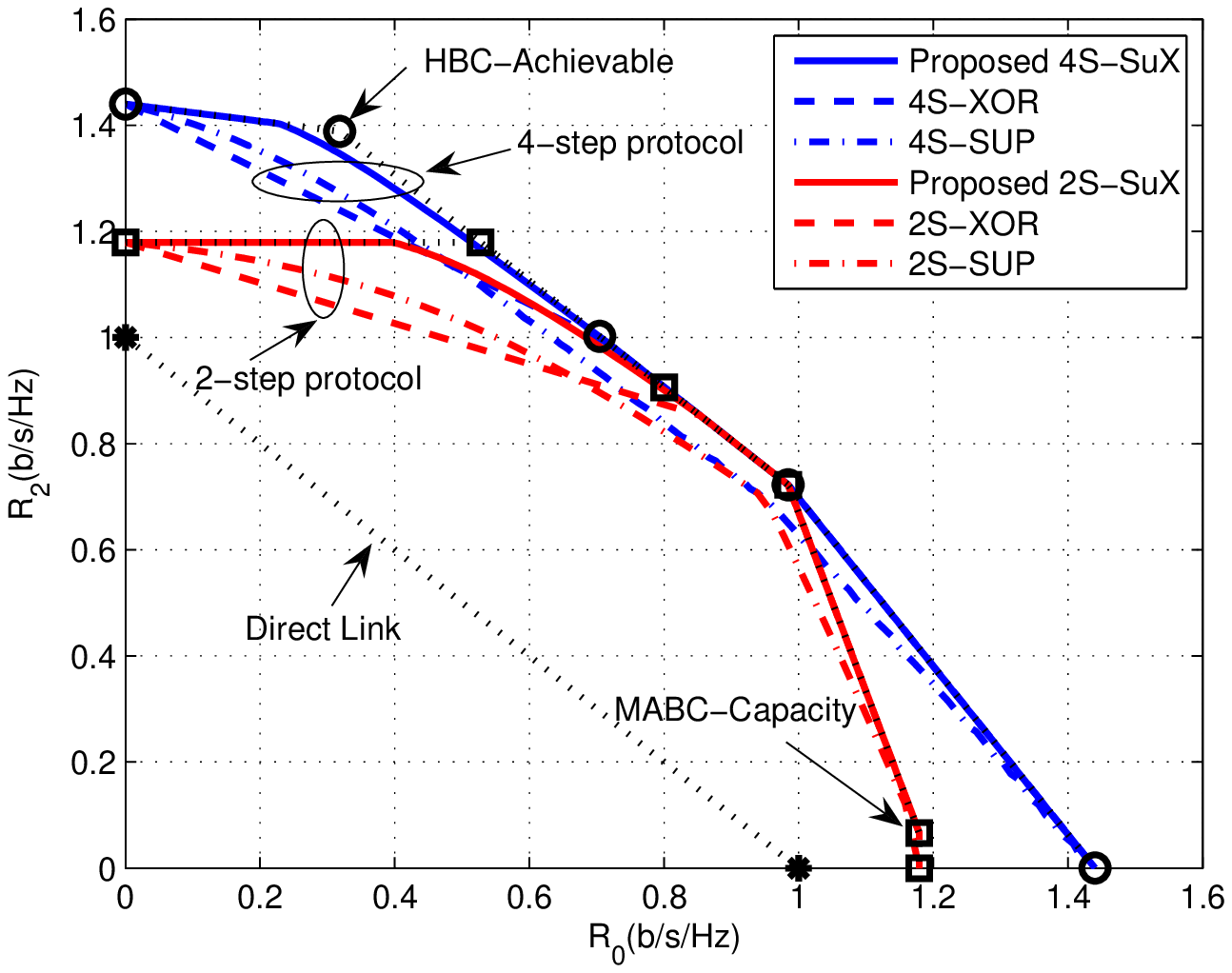}
\caption{Rate region at $(x,y)=(-0.2,0.3)$ with $P=0$dB.}
\label{fig:region_a0dB}
\end{figure}

\begin{figure}
\centering
\includegraphics[width=5in]{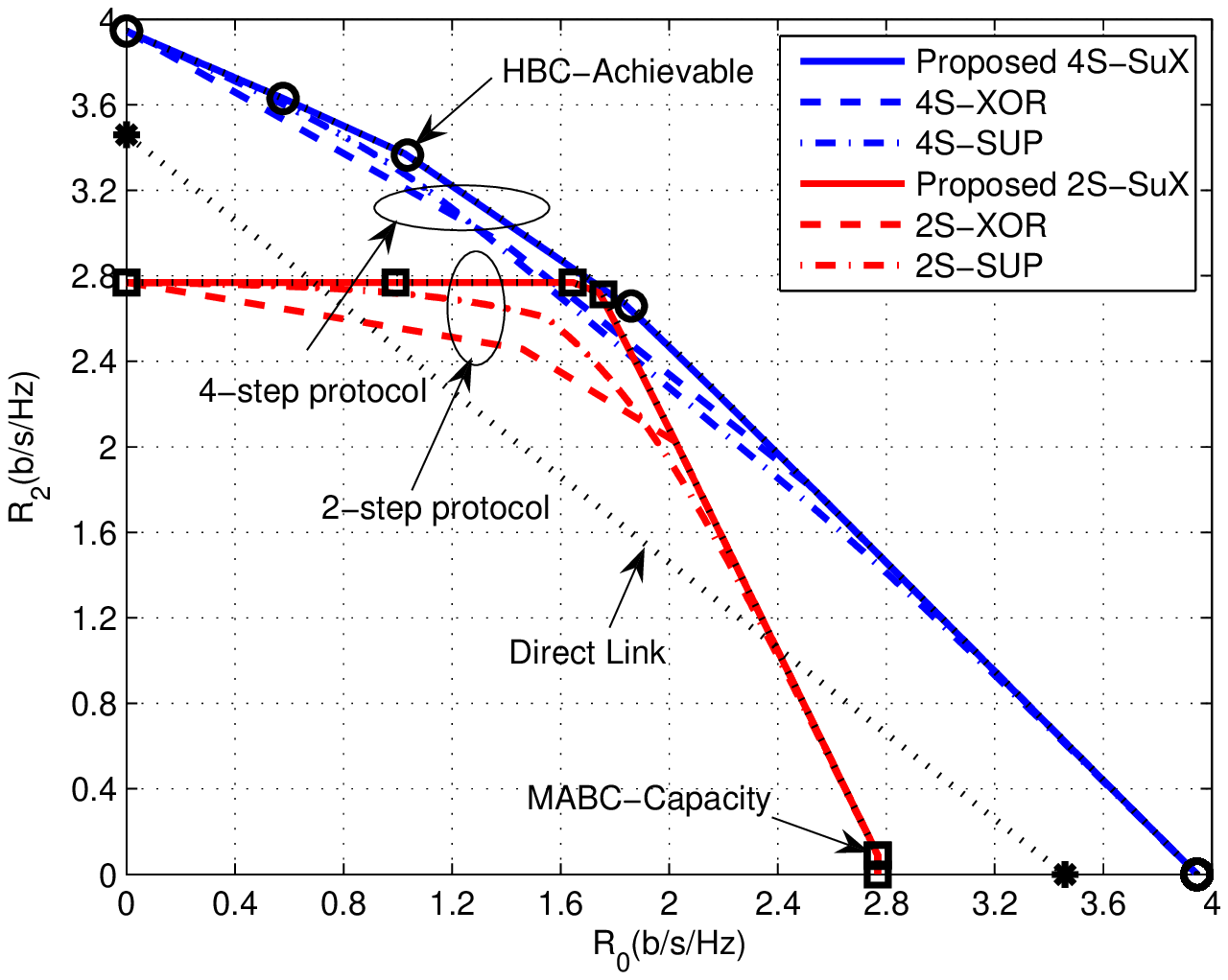}
\caption{Rate region at $(x,y)=(-0.2,0.3)$ with $P=10$dB.}
\label{fig:region_a10dB}
\end{figure}

Figs.~\ref{fig:region_s}-\ref{fig:region_a10dB} illustrate the rate
regions of different two-way relay strategies. For comparison, the
achievable rate region of the hybrid broadcast (HBC) protocol
(4-step protocol)~\cite{IT08:Kim} and the capacity of the
multiple-access broadcast (MABC) protocol (2-step protocol) derived
in~\cite{IT08:Oechtering,CWIT07:Xie,IT08:Kim} are also shown and
denoted by markers only in the figures. From
Fig.~\ref{fig:region_s}, where the two relay channels are
symmetrical, we see that the SuX and XOR schemes are equivalent and
capacity achieving, whereas the SUP scheme is much inferior. It also
shows that 4-step schemes can achieve higher one-way rate than
2-step schemes. This is expected because 4-step schemes exploit the
direct link.

From Figs.~\ref{fig:region_a0dB}-\ref{fig:region_a10dB}, where the
two relay channels are asymmetrical, it is observed that XOR is
far from capacity-achieving and that SUP schemes becomes better
than XOR schemes if $R_2 > R_0$. On the other hand, the proposed
SuX schemes closely approach the capacity (or the best achievable
rate region) in the high SNR regime (Fig.~\ref{fig:region_a10dB}),
while there is only a minor gap in the low SNR regime
(Fig.~\ref{fig:region_a0dB}).

\subsubsection{Maximum sum-rates (MSRs) in fading channels}

The problem of maximum sum-rate is formulated as
\begin{equation}
\max\limits_{(R_0(t),R_2(t))\in \mathcal {R}(t)} R_0(t) + R_2(t)
\end{equation}
where $R_k(t), k \in \{0,2\}$, denotes the service rate of node $k$
at the $t$-th time slot, and $\mathcal {R}(t)$ stands for the rate
region with respect to the channel realization at the $t$-th time
slot.

\begin{figure}
\centering
\includegraphics[width=5in]{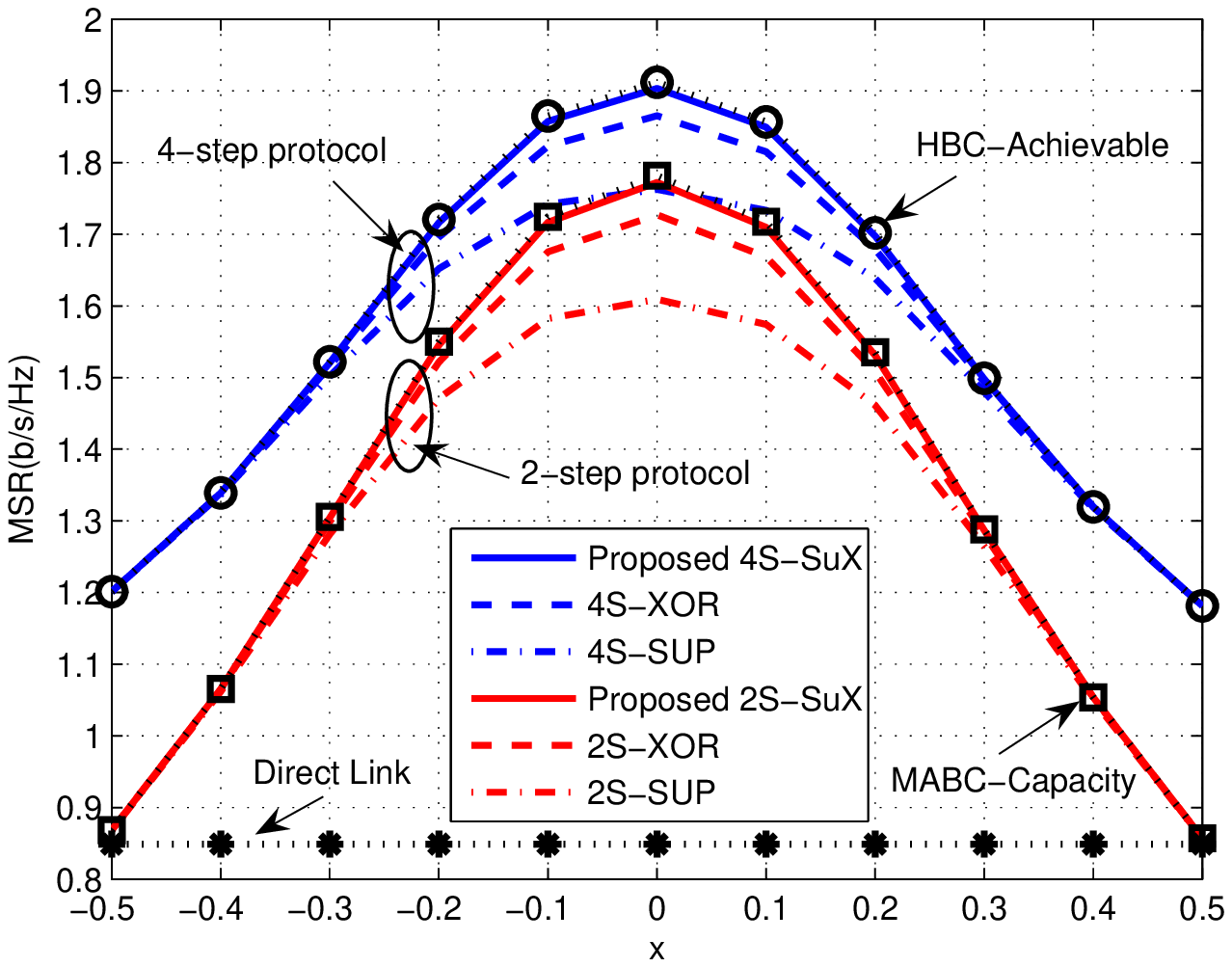}
\caption{Averaged maximum sum-rates versus relay location when $-0.5
\leq x \leq 0.5,y=0$ with $P=0$dB.} \label{fig:sumrate_p1}
\end{figure}

\begin{figure}
\centering
\includegraphics[width=5in]{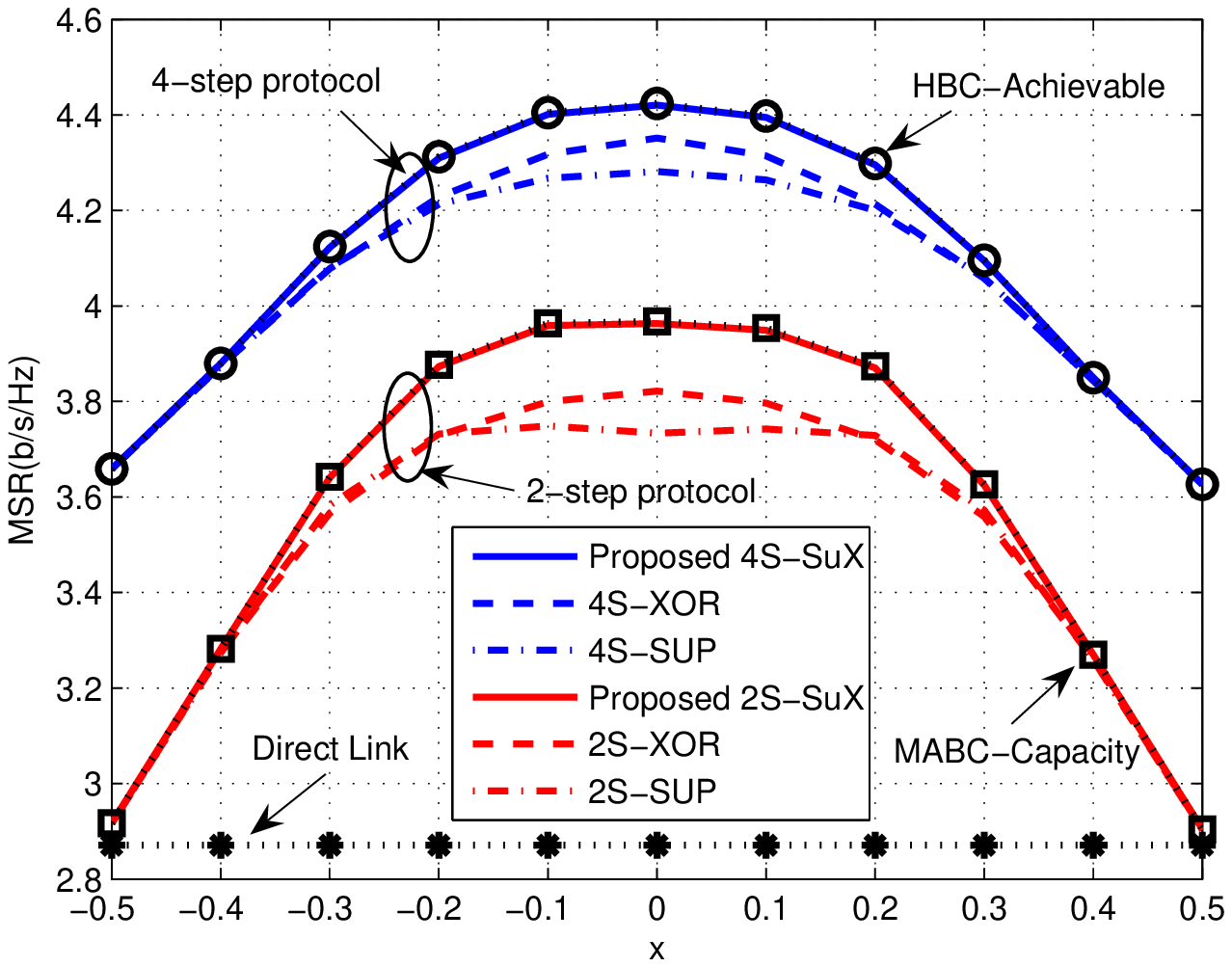}
\caption{Averaged maximum sum-rates versus relay location when $-0.5
\leq x \leq 0.5,y=0$ with $P=10$dB.} \label{fig:sumrate_p10}
\end{figure}

Figs.~\ref{fig:sumrate_p1}-\ref{fig:sumrate_p10} show the averaged
MSRs of different two-way relay strategies, when the relay node
moves alone the line between the two source nodes. It is observed
that no matter where the relay is, the proposed SuX scheme always
achieves the largest sum-rates among all the considered PLNC schemes
and approach the corresponding optimal bounds very well, especially
in the high SNR regime. Moreover, we can see that all strategies
except 2S-SUP achieve their average maximum sum-rate when the relay
node lies in the middle. As the relay node is approaching one source
node (i.e. $x$ approaches 0.5 or -0.5), the performances of the
considered PLNC schemes (XOR, SUP and SuX) converge to the optimal
bound.

\subsubsection{Service delay in fading channels}

Here, we use the same queuing mode as in~\cite{TC08:Oechtering}, in
which service rate allocation is done by a cross-layer approach by
taking into account both queue length and channel state. Note that,
Oechtering, {\it et al.} in~\cite{TC08:Oechtering} only focus on
2S-SUP strategies and consider the queue backlog versus bit arrival
rate. Chen, {\it et al.} considered delay power tradeoff
in~\cite{ICC07:Chen}, and assumed all the links have the same rate
and only the relay has buffer. We study the bit delay versus packet
arrival rate. Suppose that the packet arrival at two source nodes
follow Poisson distribution with mean $\rho$, the length of each
packet is fixed as $L$ bits. Let $Q(t-1)=[Q_0(t-1),Q_2(t-1)]$
represent the remaining bits in the queues after the $(t-1)$-th time
slot. Then, $Q(t)=[Q_0(t),Q_2(t)] = [Q_0(t-1) - R_0(t) +
A_0(t)L,Q_2(t-1) - R_2(t) + A_2(t)L]$, where $R_k(t)$,$A_k(t)$ ($k
\in \{0,2\}$) denote the service rates and packet arrival rates of
node $k$ in $t$-th time slot. The rate allocation problem is
formulated as
\begin{equation}
\max\limits_{\stackrel{(R_0(t),R_2(t))\in \mathcal {R}(t)}{R_0(t)
\leq Q_0(t-1), R_2(t) \leq Q_2(t-1)}} Q_0(t-1)R_0(t) +
Q_2(t-1)R_2(t)
\end{equation}

\begin{figure}
\centering
\includegraphics[width=5in]{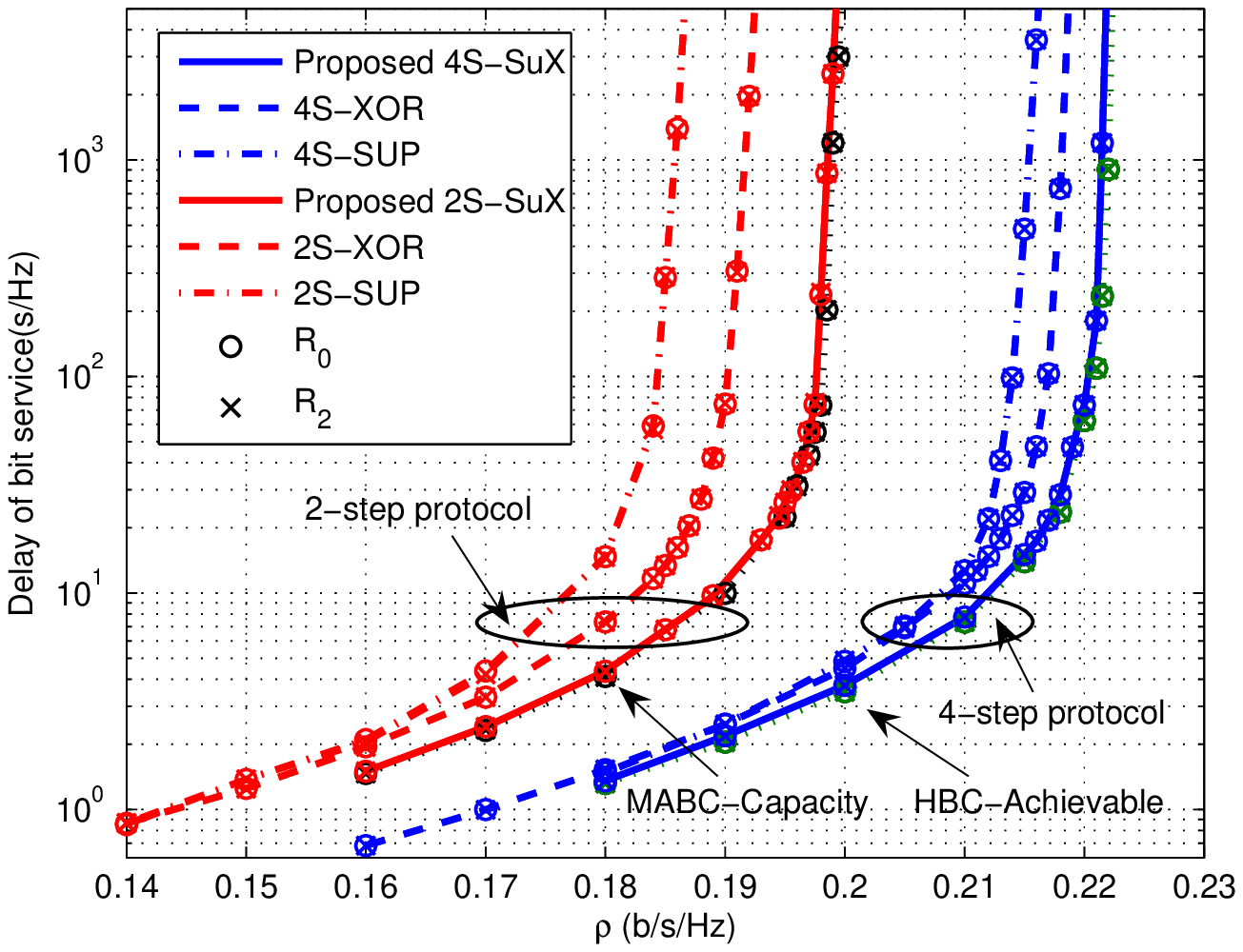}
\caption{System service delay versus packet arrival rate:
$(x,y)=(0,0)$, $P=10$dB.} \label{fig:queue}
\end{figure}

Fig.~\ref{fig:queue} shows the system delay based on the above
weighted sum-rate maximization with $L=10$. It can be clearly seen
that the proposed SuX scheme always outperforms the other two PLNC
schemes and approach the corresponding best achievable bound, no
matter which time control protocol is applied.

\section{Analysis of Performance Gap}

The numerical results in the previous section demonstrate that the
rate region achieved by proposed superimposed XOR closely approach
the capacity bound. In this section, we shall analytically quantify
the performance gap and show that it indeed approaches zero when SNR
is large.

Note that the proposed scheme only concerns the information
processing at the relay and destination during the broadcast phase,
no matter which time protocol is adopted. Hence, it suffices to
analyze the gap on the broadcast rate region. In what follows we
first deduce the boundary of the achievable broadcast rate region in
an analytical expression. Then we characterize an upper bound on the
capacity gap.

\begin{figure}
\centering
\includegraphics[width=5in]{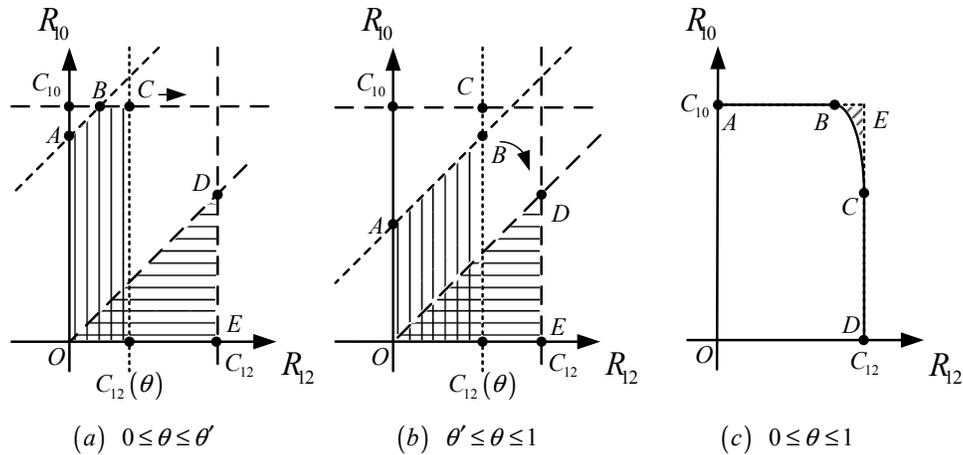}
\caption{Rate region of superimposed XOR: shadowed area in (a) or
(b) is the rate region at a fixed $\theta$; solid lines and dashed
lines in (c) are the boundary of superimposed XOR and capacity
bound, respectively.}
 \label{fig:line_region}
\end{figure}

Let $R_{10}$ and $R_{12}$ denote the data rates of the information
flows $1 \rightarrow 0$ and $1 \rightarrow 2$, respectively, in the
broadcast phase. From the proof of Theorem 1, it is seen clearly
that, for any given power allocation parameter $\theta$, the rate
pair $(R_{10}, R_{12})$ must satisfy the following linear
inequalities:
\begin{eqnarray*}
\label{eq:BCphase} R_{10} &\leq & C_{10}, \\
R_{12} &\leq & C_{12}(\theta), \\
R_{10}-R_{12} &\leq & C_{10}(1-\theta).
\end{eqnarray*}
for $R_{10} \ge R_{12}$, and satisfy
\begin{eqnarray*}
  R_{12} &\le & C_{12}
\end{eqnarray*}
for $R_{10}\le R_{12}$.
Graphing the feasible set and considering that $R_{10} \ge 0$ and
$R_{12} \ge 0$, we obtain the rate region as sketched in
Fig.~\ref{fig:line_region}. Here, we have
\begin{equation}\label{eqn:theta1}
 \theta' = \left[1 + \frac{1}{\gamma_{10}}  - \frac{1}{\gamma_{12}}
\right]^+,
\end{equation}
with $[x]^+ = x$ if $x \geq 0$ and $[x]^+ = 0$ otherwise. This power
control value satisfies $C_{10} = C_{12}(\theta') +
C_{10}(1-\theta')$.

When $0 \leq \theta \leq \theta'$ or equivalently $ C_{10} \le
C_{12}(\theta) + C_{10}(1-\theta)$, the rate region is plotted in
Fig.~\ref{fig:line_region}(a), with A= $(0,C_{10}(1-\theta))$, B =
$(C_{10}-C_{10}(1-\theta),C_{10})$, C = $(C_{12}(\theta),C_{10})$
and D = $(C_{12},C_{12})$. On the other hand, when $\theta' \leq
\theta \leq 1$, or equivalently, $C_{10} \ge
C_{12}(\theta)+C_{10}(1-\theta)$, the rate region is in
Fig.~\ref{fig:line_region}(b), where A= $(0,C_{10}(1-\theta))$, B =
$(C_{12}(\theta),C_{12}(\theta)+C_{10}(1-\theta))$ and D =
$(C_{12},C_{12})$.

Then, by varying $\theta$ from 0 to 1, the overall broadcast rate
region is obtained as the union of the above feasible sets, which is
given in Fig.~\ref{fig:line_region}(c). The coordinates of the
boundary in the segment $\widetilde{{\rm {BC}}}$ can be
characterized as:
\begin{equation*}
\begin{array}{rl}
  {\rm B}: & (C_{12}(\theta'), C_{10} ) \\
  {\rm C}: & (C_{12}, C_{12}) \\
  {\widetilde{\rm{BC}}}: & (C_{12}(\theta), C_{12}(\theta) +
  C_{10}(1-\theta)), \forall \theta \in [\theta', 1].
\end{array}
\end{equation*}
In Fig.~\ref{fig:line_region}(c), the capacity bound of two-way
relaying in the broadcast phase is also shown for comparison. It is
given by the rectangle characterized by $R_{10} \le C_{10}$ and
$R_{12} \le C_{12}$ \cite{IT08:Oechtering}.

We now define the gap between the rate region of the superimposed
XOR and the capacity bound as the area of the shadowed region as
depicted in Fig.~\ref{fig:line_region}(c)($\widetilde{BC}$,
$\overline{CE}$ and $\overline{EB}$), denoted as $\Delta$. Thus,
\begin{eqnarray*}
\Delta &=&\int\limits_{\theta'}^{1}\Big\{C_{10}-\big[C_{12}(\theta)+C_{10}(1-\theta)\big]\Big\}\emph{d}C_{12}(\theta) \nonumber\\
&=&\int\limits_{\theta'}^{1}\frac{\gamma_{12}}{(1+\theta\gamma_{12})\ln2}\log_2\frac{1+\gamma_{10}}{(1+\theta\gamma_{12})\big[1+(1-\theta)\gamma_{10}\big]}\emph{d}\theta\\
\end{eqnarray*}
However, computing the exact value of $\Delta$ is involved. It can
be verified easily that the gap is upper-bounded by the area of the
triangular formed by $\overline{BC}$, $\overline{CE}$ and
$\overline{EB}$. Namely,
\begin{eqnarray}
\Delta
& \leq &\frac{1}{2}(C_{10}-C_{12})\Big[C_{12}-C_{12}(\theta')\Big]\\
& =
&\frac{1}{2}\log_2\frac{1+\gamma_{10}}{1+\gamma_{12}}\log_2\frac{\gamma_{10}(1+\gamma_{12})}{\gamma_{12}(1+\gamma_{10})}
\end{eqnarray}

It is seen that when both $\gamma_{10}$ and $\gamma_{12}$ are much
large than one ($\gamma_{10} \geq \gamma_{12} \gg 1$), then the term
$\log_2(\gamma_{10}(1+\gamma_{12})/(\gamma_{12}(1+\gamma_{10})))$
approaches zero. Therefore, the upper bound approaches zero.
Finally, we conclude that the proposed scheme is capacity
approaching for larger SNR.

\begin{figure}
\centering
\includegraphics[width=5in]{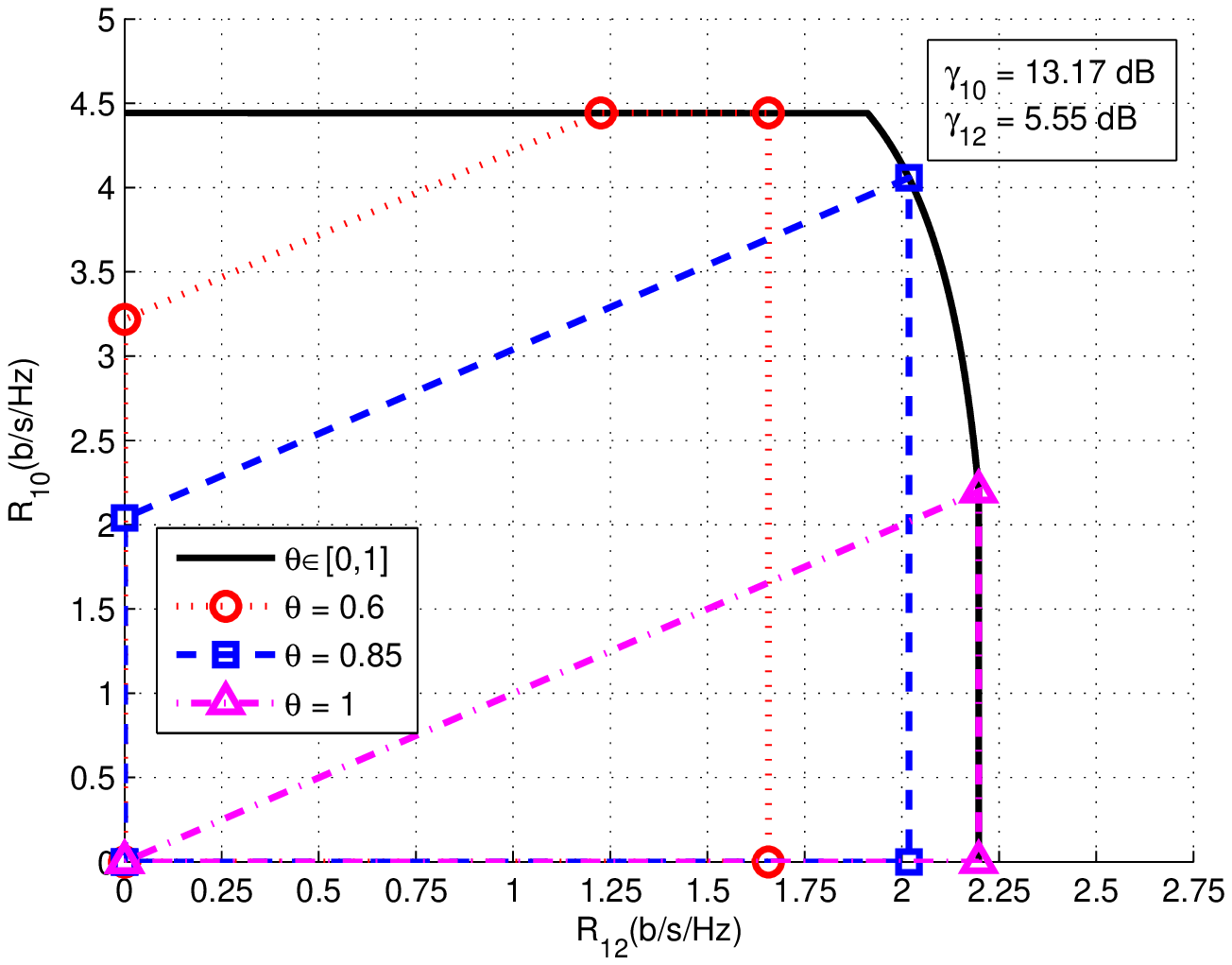}
\caption{Rate regions of superimposed XOR at fixed $\theta$ and the
boundary for $\theta \in [0,1]$ in broadcast phase: $(x,y)=( -0.2,
0.3)$, $P = 2$dB.}
 \label{fig:BCregion_example}
\end{figure}

In the following we show some numerical examples for further
illustration. Fig.~\ref{fig:BCregion_example} demonstrates the rate
regions of superimposed XOR for different $\theta$ as well as the
overall broadcast rate region by letting $\theta$ take all possible
values in $[0,1]$. In this example, we have $\gamma_{10}=13.17$ dB
and $\gamma_{12}=5.55$ dB. The corresponding $\theta'$ is $0.77$.

\begin{figure}
\centering
\includegraphics[width=5in]{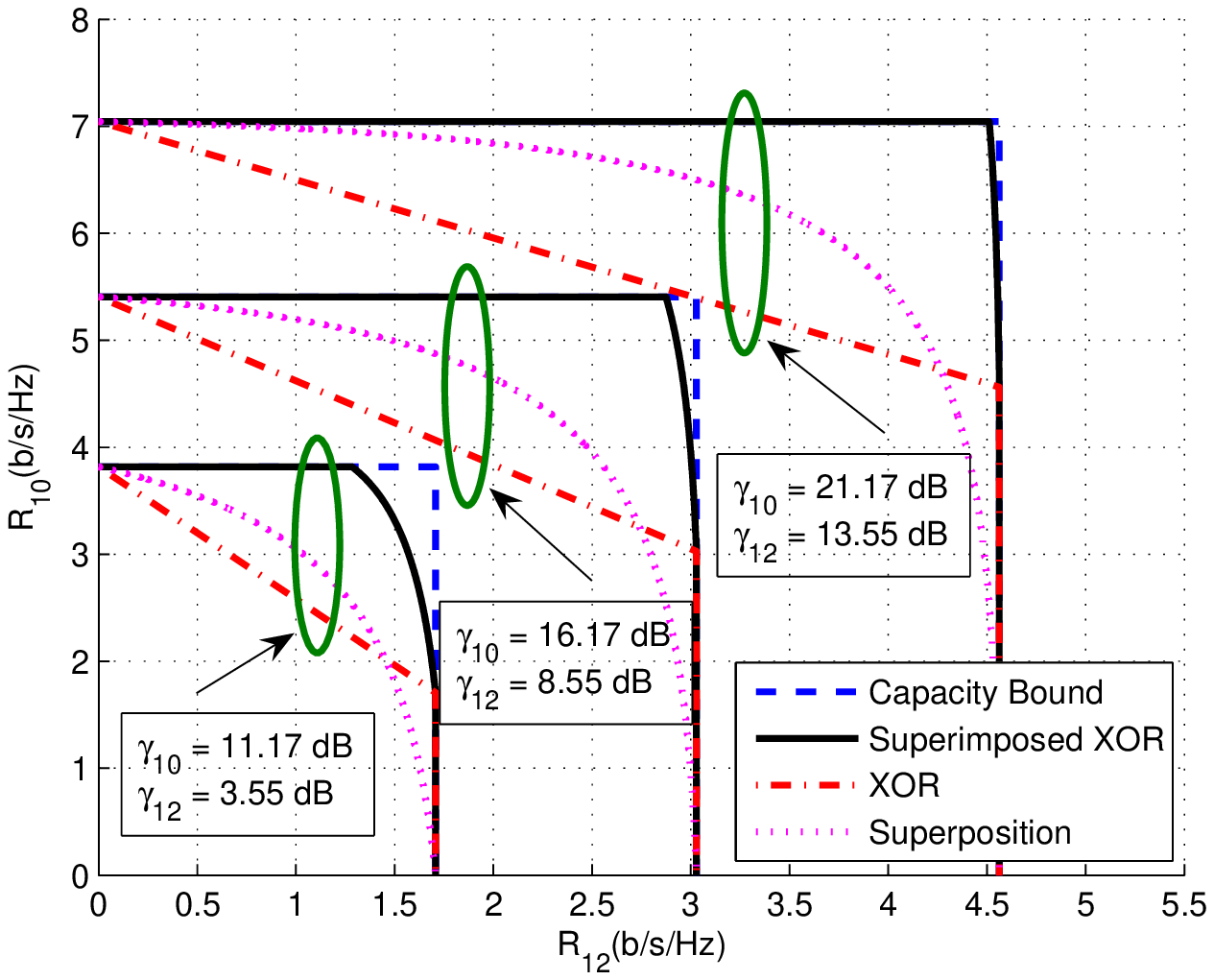}
\caption{Comparisons of rate regions between the superimposed XOR
and capacity bound in broadcast phase: $(x,y) = (-0.2, 0.3)$, $P=
\{0,5,10\}$ dB.}
 \label{fig:BCregion_compare}
\end{figure}

Fig.~\ref{fig:BCregion_compare} shows the broadcast rate regions of
superimposed XOR compared to the capacity bound and the rate regions
obtained by conventional XOR and superposition at different SNR. It
can be seen that the proposed superimposed XOR outperforms the other
two schemes, and the gap to the capacity bound vanishes as SNR
increases.

\section{Conclusion}

In this research, we proposed superimposed XOR, a novel PLNC scheme
for two-way relay communications. It takes into consideration the
asymmetry in both channel gain and bidirectional information length
during the broadcast phase. In specific, when the receiver channel
quality of the longer packet is worse than that of the shorter
packet, it reduces to the conventional XOR. Otherwise, it combines
the extra information bits from the longer packet with the XORed
bits using superposition coding. We characterized its achievable
rate region over the Gaussian channel when applied together with the
4-step and 2-step transmission protocols.
We also demonstrate its average maximum sum-rate and service delay
performance over fading channels.
Numerical results showed that proposed PLNC scheme achieves a larger
rate region than XOR and superposition when the broadcast channels
are asymmetric
and performs much better over fading channels.
Moreover, at the high SNR region, it approaches the capacity bound.
We also explicitly proved this capacity approaching behavior by
deriving the analytical expressions of the boundary of the broadcast
rate region.

\bibliographystyle{IEEEtran}
\bibliography{IEEEabrv,reference}

\begin{thebibliography}{10}
\providecommand{\url}[1]{#1}
\csname url@rmstyle\endcsname
\providecommand{\newblock}{\relax}
\providecommand{\bibinfo}[2]{#2}
\providecommand\BIBentrySTDinterwordspacing{\spaceskip=0pt\relax}
\providecommand\BIBentryALTinterwordstretchfactor{4}
\providecommand\BIBentryALTinterwordspacing{\spaceskip=\fontdimen2\font plus
\BIBentryALTinterwordstretchfactor\fontdimen3\font minus
  \fontdimen4\font\relax}
\providecommand\BIBforeignlanguage[2]{{%
\expandafter\ifx\csname l@#1\endcsname\relax
\typeout{** WARNING: IEEEtran.bst: No hyphenation pattern has been}%
\typeout{** loaded for the language `#1'. Using the pattern for}%
\typeout{** the default language instead.}%
\else
\language=\csname l@#1\endcsname
\fi
#2}}

\bibitem{GC09:We}
J.~Liu, M.~Tao, Y.~Xu, and X.~Wang, ``Superimposed xor: A new physical layer
  network coding scheme for two-way relay channels,'' in \emph{Proc. IEEE
  Global Telecomm. Conf. (GLOBECOM)}, Nov. 2009.

\bibitem{AAP71:Meulen}
E.~C. V.~D. Meulen, ``Three-terminal communication channels,'' in
  \emph{Advances in Applied Probability}, vol.~3.\hskip 1em plus 0.5em minus
  0.4em\relax Applied Probability Trust, 1971, pp. 120--154.

\bibitem{IT79:Cover}
T.~M. Cover and A.~E. Gamal, ``Capacity theorems for the relay channel,''
  \emph{IEEE Trans. Inform. Theory}, vol.~25, no.~5, pp. 572--584, Sept. 1979.

\bibitem{CISS05:Wu}
Y.~Wu, P.~A. Chou, and S.-Y. Kung, ``Information exchange in wireless networks
  with network coding and physical-layer broadcast,'' in \emph{Proc. 39th Conf.
  Inform. Sciences Systems (CISS)}, Mar. 2005.

\bibitem{ADHOC05:Larsson}
P.~Larsson, N.~Johansson, and K.-E. Sunell, ``Coded bi-directional relaying,''
  in \emph{Proc. 5th Scandinavian Workshop Ad Hoc Networks (ADHOC)}, May 2005.

\bibitem{VTC06:Popovski}
P.~Popovski and H.~Yomo, ``Bi-directional amplification of throughput in a
  wireless multi-hop network,'' in \emph{Proc. IEEE Veh. Tech. Conf. (VTC)},
  May 2006, pp. 588--593.

\bibitem{ICC06:Hausl}
C.~Hausl and J.~Hagenauer, ``Iterative network and channel decoding for the
  two-way relay channel,'' in \emph{Proc. IEEE Int. Conf. Comm. (ICC)}, Jun.
  2006, pp. 1568--1573.

\bibitem{AMC06:Zhang}
S.~Zhang, S.~C. Liew, and P.~P. Lam, ``Physical-layer network coding,'' in
  \emph{Proc. ACM 15th Annual Int. Conf. Mobile Computing Networking
  (MobiCom)}, Sept. 2006, pp. 358--365.

\bibitem{ASC06:Katti}
S.~Katti, H.~Rahul, W.~Hu, D.~Katabi, M.~M. edard, and J.~Crowcroft, ``{XOR}s
  in the air: Practical wireless network coding,'' in \emph{Proc. ACM SIGCOMM},
  Sept. 2006, pp. 243--254.

\bibitem{JSAC07:Rankov}
B.~Rankov and A.~Wittneben, ``Spectral efficient protocols for half-duplex
  fading relay channels,'' \emph{IEEE Jour. Selec. Areas. Comm.}, vol.~25,
  no.~2, pp. 379--389, Feb. 2007.

\bibitem{IT00:Ahlswede}
R.~Ahlswede, N.~Cai, S.-Y.~R. Li, and R.~W. Yeung, ``Network information
  flow,'' \emph{IEEE Trans. Inform. Theory}, vol.~46, no.~4, pp. 1204--1216,
  July 2000.

\bibitem{IT08:Oechtering}
T.~J. Oechtering, C.~Schnurr, I.~Bjelakovic, and H.~Boche, ``Broadcast capacity
  region of two-phase bidirectional relaying,'' \emph{IEEE Trans. Inform.
  Theory}, vol.~54, no.~1, pp. 454--458, Jan. 2008.

\bibitem{CWIT07:Xie}
L.~L. Xie, ``Network coding and random binning for multi-user channels,'' in
  \emph{Proc. 10th Canadian Workshop Inf. Theory}, June 2007, pp. 85--88.

\bibitem{IT08:Kim}
S.~J. Kim, P.~Mitran, and V.~Tarokh, ``Performance bounds for bi-directional
  coded cooperation protocols,'' \emph{IEEE Trans. Inform. Theory}, vol.~54,
  no.~11, pp. 5235--5241, Nov. 2008.

\bibitem{ISS08:Kim}
S.~J. Kim, N.~Devroye, P.~Mitran, and V.~Tarokh, ``Comparison of bi-directional
  relaying protocols,'' in \emph{Proc. IEEE Sarnoff Symposium}, April 2008, pp.
  1--5.

\bibitem{ICC07:Popovski}
P.~Popovski and H.~Yomo, ``Physical network coding in two-way wireless relay
  channels,'' in \emph{Proc. IEEE Int. Conf. Comm. (ICC)}, June 2007, pp.
  707--712.

\bibitem{ICC08:Liu}
C.~H. Liu and F.~Xue, ``Network coding for two-way relaying: rate region, sum
  rate and opportunistic scheduling,'' in \emph{Proc. IEEE Int. Conf. Comm.
  (ICC)}, May 2008, pp. 1044--1049.

\bibitem{CL08:Oechtering}
T.~J. Oechtering and H.~Boche, ``Optimal time-division for bi-directional
  relaying using superposition encoding,'' \emph{IEEE Comm. Letters}, vol.~12,
  no.~4, pp. 265--267, April 2008.

\bibitem{CC09:We}
J.~Liu, M.~Tao, and Y.~Xu, ``Rate regions of a two-way gaussian relay
  channel,'' in \emph{Proc. Int. Conf. Comm. Networking China (ChinaCom)}, Aug.
  2009.

\bibitem{GC08:Liu}
C.~H. Liu and A.~Arapostathis, ``Joint network coding and superposition coding
  for multi-user information exchange in wireless relaying networks,'' in
  \emph{Proc. IEEE Global Telecomm. Conf. (GLOBECOM)}, Nov. 2008, pp. 1--6.

\bibitem{ICC06:Chen}
W.~Chen, K.~B. Letaief, and Z.~Cao, ``A cross layer method for interference
  cancellation and network coding in wireless networks,'' in \emph{Proc. IEEE
  Int. Conf. Comm. (ICC)}, May 2006, pp. 3693--3698.

\bibitem{IT06:Anders}
A.~H.-Madsen, ``Capacity bounds for cooperative diversity,'' \emph{IEEE Trans.
  Inform. Theory}, vol.~52, no.~4, pp. 1522--1544, April 2006.

\bibitem{TC08:Oechtering}
T.~J. Oechtering and H.~Boche, ``Stability region of an optimized
  bi-directional regenerative half-duplex relaying protocol,'' \emph{IEEE
  Trans. Comm.}, vol.~56, no.~9, pp. 1519--1529, Sep. 2008.

\bibitem{ICC07:Chen}
W.~Chen, K.~B. Letaief, and Z.~Cao, ``Opportunistic network coding for wireless
  networks,'' in \emph{Proc. IEEE Int. Conf. Comm. (ICC)}, 2007, pp.
  4634--4639.

\end{thebibliography}

\newpage


%

\end{document}